\documentclass[runningheads]{llncs}
\usepackage[T1]{fontenc}
\usepackage{graphicx}
\usepackage{booktabs}
\usepackage[misc]{ifsym}
\newcommand{\corr}{(\Letter)}
\usepackage{microtype}
\usepackage{graphicx}
\usepackage{subfigure}
\usepackage{booktabs} % for professional tables
\usepackage{tikz,pgfplots}
\usepackage{tikz}
\usepackage{paralist}
\usepackage{adjustbox}
\usepackage{graphicx}
\usepackage{enumitem}
\usepackage{amsmath}
\usepackage{tcolorbox}
\usepackage{multirow}
\usepackage{hyperref}

\usepackage{mathtools}
\usepackage{amssymb}

\usepackage[capitalize,noabbrev]{cleveref}

\usepackage[textsize=tiny]{todonotes}
\newcommand{\flhhe}{\texttt{FLHHE}}
\newcommand{\simpavg}{\texttt{SimpAvg}}
\begin{document}

\title{A Privacy-Centric Approach: Scalable and Secure Federated Learning Enabled by Hybrid Homomorphic Encryption}

\titlerunning{A Privacy-Centric Approach}
% If the full title of your paper is short enough to also fit in the running head, you can omit the abbreviated paper title here. You can check as follows: if you comment out the \titlerunning line, something will appear in the header of all odd-numbered pages of your PDF from page 3 onward. This something is either the full title (in which case all is well), or the error message "Title Suppressed Due to Excessive Length". If this error message appears, you're going to want to provide an abbreviated title within the \titlerunning command, because if you won't do it, Springer will do it for you.

%N.B.: Author information (both in the \author{} and \authorrunning{} command) should only be present in the Camera-Ready Version of your paper. The version that you initially submit for review, ought to be double-blind. So, when initially submitting your paper, use:
% \author{Author information scrubbed for double-blind reviewing}
\author{Khoa Nguyen \and
Tanveer Khan \corr \and
Hossein Abdinasibfar \and
Antonis Michalas}
% You may leave out the orcidID information, if you want to.
% Use \corr to indicate the corresponding author. Note the spacing around the \corr command. Only one author can be the corresponding author.

%N.B.: comment out the \authorrunning{} command for the double-blind version of your paper submitted for review. Later, if your paper is accepted, use the command for the Camera-Ready Version.
\authorrunning{K. Nguyen et al.}
% If there are two authors, write 'A.L. Benjamin and C.C. Broadus Jr.'
% If there are more than two authors, '[...] et al.' is used.

\institute{Tampere University, Tampere, Finland \\ \email{\{khoa.nguyen, tanveer.khan, hossein.abdinasibfar, antonios.michalas\}@tuni.fi}
% \and
% Fictional West Coast University, Long Beach CA 90840, USA \email{ccb@fwcu.fake}
% \and
% Secondary European Affiliation, Tiergartenstr. 17, 69121 Heidelberg, Germany
% \email{lncs@springer.com}
}

\maketitle              % typeset the header of the contribution

\begin{abstract}
Federated Learning (FL) enables collaborative model training without sharing raw data, making it a promising approach for privacy-sensitive domains. Despite its potential, FL faces significant challenges, particularly in terms of communication overhead and data privacy. Privacy-preserving Techniques (PPTs) such as Homomorphic Encryption (HE) have been used to mitigate these concerns. However, these techniques introduce substantial computational and communication costs, limiting their practical deployment. In this work, we explore how Hybrid Homomorphic Encryption (HHE), a cryptographic protocol that combines symmetric encryption with HE, can be effectively integrated with FL to address both communication and privacy challenges, paving the way for scalable and secure decentralized learning system. %To our knowledge, we are the first one to attempt combining HHE with FL.

\end{abstract}

\section{Introduction}
\label{sec: introduction}
Machine learning (ML) has increasingly become one of the most impactful fields of data science in recent years, allowing various users to classify and make predictions based on multidimensional data. ML applications range from recommendation systems~\cite{pazzani2007content} to medical diagnostics~\cite{fatima2017survey}. One of the primary metrics for ML is the accuracy of the prediction or classification results. However, to achieve this, the results should be accompanied by a large amount of high-quality training data that requires the collaboration of several organizations. New regulations such as the General Data Protection Regulations (GDPR) forbid the sharing and processing of sensitive data without the consent of the data subject. This is a key limitation especially when the training data contains sensitive information and therefore poses security threats. For example, to develop a breast cancer detection model from MRI scans, different hospitals can share their data to develop a collaborative ML model. However, sharing private patient information to a central server can reveal sensitive information to the public with several repercussions. Therefore, it has become crucial to protect data privacy, confidentiality, and profit sharing while obtaining data from other organizations. One solution to this problem is employing Federated Learning (FL). 

FL enables collaborative model training across edge devices without sharing raw data, making it suitable for domains like healthcare and autonomous transportation~\cite{konevcny2016federated}. A key challenge in FL is ensuring data privacy while managing communication overhead and data heterogeneity. Although FL keeps data local, Adversaries ($\mathcal{ADV}$) can still infer private information~\cite{li2021survey}, which necessitates Privacy-preserving Techniques (PPTs).

Among PPTs, Homomorphic Encryption (HE))~\cite{tian2022sphinx,khan2024wildest} and Secure Multi-party Computation (SMPC)~\cite{wagh2019securenn,ryffel2020AriaNN,khan2024make} are widely used. HE allows computation on encrypted data, enabling privacy without compromising utility. However, HE introduces high computational and communication costs, and large ciphertext sizes making real-world deployment difficult~\cite{khan2023love,dobraunig2021pasta,frimpong2024guardml,nguyen2023split,khan2025split}. To address this Hybrid Homomorphic Encryption (HHE) combines symmetric encryption with HE, significantly reducing ciphertext size and communication cost~\cite{dobraunig2021pasta,bakas2019modern} . Users encrypt data with a symmetric key and then encrypt the key using HE. The server uses the encrypted key to process the data securely. HHE thus enables practical deployment of HE on consumer devices by offloading complex computations to the servers.  Researchers have deployed HE-friendly symmetric ciphers (e.g, HERA/Rubato~\cite{cho2021transciphering,ha2022rubato}, Elisabeth~\cite{cosseron2022towards} and PASTA~\cite{dobraunig2021pasta}) to support this approach. HHE effectively mitigates both communication and privacy challenges in FL. In this paper, we aim to explore the effectiveness and feasibility of combining FL with HHE. Specifically, we focus on:

\begin{tcolorbox}[colback=gray!5!white, colframe=gray!75!black, title=Research Questions] % Example customizations
\begin{itemize}[left=0pt, nosep]
    \item Can we combine FL and HHE to achieve shared trained models without leaking data privacy?
    \item Does HHE degrade accuracy of the aggregated model compared to plaintext and HE version?
    \item What are the trade-offs we have to make? 
    \begin{itemize}[left=0pt]
        \item How much computation and communication does each client save?
        \item How much more computation is needed on the server to compensate?
    \end{itemize}
\end{itemize}
\end{tcolorbox}

\section{Related Works}
\label{sec:relatedwork}

% \noindent\textbf{Federated Learning:}
FL trains ML models on distributed devices by aggregating local model updates instead of local data. While FL ensures that local raw data do not leave their original locations, it remains vulnerable to eavesdroppers and malicious FL servers that might exploit local plaintext models to reconstruct sensitive training data~\cite{fredrikson2015model,zhu2019deep,geiping2020inverting}. One measure to protect against privacy breaches is differential privacy in which noise is added to the training updates by each client. However, while this paradigm protects private information, it comes at a utility trade-off and can lead to fewer performance models, as demonstrated recently. Another PPT which could be used to address the privacy issue of FL is HE. HE can protect against malicious server eavesdroppers while maintaining model performance by encrypting the weight before sending them to the central server. In HE based FL, the goal is to ensure that no client reveals its model updates during aggregation. Several approaches~\cite{park2022privacy,yang2019federated} have been proposed to achieve this, including the use of a partially HE scheme~\cite{fang2021privacy}. With HE, gradient aggregation can be performed on ciphertexts without decrypting them in advance. For example, Mandal and Gong~\cite{mandal2019privfl} created robust and secure training protocols for federated regression models using HE. Park and Lim~\cite{park2022privacy} propose employing an HE scheme that enables the centralized server to aggregate encrypted local model parameters without decryption and allows each client to use a different HE key in the same FL system using a distributed cryptosystem. In another work, Madi \textit{ et al.}~\cite{madi2021secure} propose an FL framework that is secure against both confidentiality and integrity threats from the aggregation server by employing HE and verifiable computing. Stripelis \textit{et al.,},~\cite{stripelis2021secure}, propose a secure FL framework using FHE with the CKKS scheme. Testing on the MRI dataset shows no performance loss between encrypted and non-encrypted models. Shi \textit{et al.,} in paper~\cite{shi2023privacy} propose a privacy-preserving scheme combining secret sharing and encryption to protect local parameters and resist collusion in FL. The authors also claim that the proposed protocol supports client dropouts, keyless aggregation, and simple interactions. In short, the use of HE schemes in the implementation of PPML continues to garner attention. Gentry's work~\cite{gentry2009fully} revolutionized the field of HE and paved the way for multiple modern schemes, such as TFHE~\cite{chillotti2016faster}, BFV~\cite{fan2012somewhat}, and CKKS~\cite{cheon2017homomorphic} in PPML applications. 

BFV allows a limited number of operations over integer ciphertext while CKKS~\cite{cheon2017homomorphic} extends this to floating-point data. TFHE~\cite{chillotti2016faster} improves upon bootstrapping and bitwise operations on binary data without batching~\cite{clet2021bfv}. BFV and CKKS allow batching for faster ML computation, while TFHE uses look-up tables and packing. Each scheme handles non-linear activation functions in ML differently: TFHE uses look-up tables, and BFV/CKKS uses polynomial approximation~\cite{lee2022privacy,khan2023learning}. 
Although guaranteeing up to a high degree of privacy, a major downside of these algorithms is that they demand significant computational resources. HE schemes have large ciphertexts and high computational complexity, limiting their use. To address these issues, HHE ~\cite{dobraunig2021pasta,bakas2019modern} combines symmetric ciphers, making it more practical for users.

The first approaches to designing HHE schemes relied primarily on existing and well-established symmetric ciphers such as AES~\cite{gentry2012homomorphic}. However, these were not suitable for HHE schemes due to their large multiplicative depth~\cite{dobraunig2021pasta}. As a result of this limitation, research in the field of HHE has mainly focused on the design of compatible symmetric ciphers with different optimization criteria~\cite{bakas2022symmetrical}, such as eliminating the ciphertext expansion~\cite{canteaut2018stream} or using filter permutators~\cite{meaux2019improved}. However, to date, HHE has seen limited practical use~\cite{bakas2022symmetrical} in real-world applications, and only a handful of works exist in the field of PPML. To the best of our knowledge, the main HHE schemes currently are HERA~\cite{cho2021transciphering}, Elisabeth~\cite{cosseron2022towards} and PASTA~\cite{dobraunig2021pasta}. The authors of HERA also proposed Rubato~\cite{ha2022rubato}; however, the specifications remain largely the same as in HERA. These proposed approaches have different specifications and apply to different use cases. %We provide an overview of these approaches in \autoref{tab:comp-hhe-schemes}. 
HERA~\cite{cho2021transciphering} is a stream cipher based on the CKKS HE scheme and allows computations on floating-point data types. In comparison, Elisabeth~\cite{cosseron2022towards} utilizes the TFHE scheme, is defined over $\mathbb{Z}_\mathsf{q}$, where $\mathsf{q} = 2^4$, and stores up to 4 bits of data. PASTA~\cite{dobraunig2021pasta} utilizes the BFV scheme for integer data types. HERA and PASTA are defined over $\mathbb{Z}_\mathsf{q}$, where $\mathsf{q} = 2^{16} + 1$, and store up to 16-bit data. Each scheme achieves the same security level of~128-bits. Additionally, HERA also provides tests for a security level of 80 bits. As HERA allows computations on floating-point data types, it does not require quantization on certain inputs. Elisabeth and PASTA, on the other hand, require quantization to operate on floating-point numbers, which introduces a rounding error, which can reduce the accuracy of certain applications, i.e. ML.

\section{Preliminaries}
\label{sec: preliminaries}

\subsection{Federated Learning}
\label{subsec:federated-learning}

The concept of FL was first introduced by Google AI Blog\footnote{\url{{https://ai.googleblog.com/2017/04/federated-learning-collaborative.html}}} to protect data privacy. It is a distributed ML approach, where multiple users collaboratively train a model~\cite{mcmahan2023communicationefficientlearningdeepnetworks}. In this technique, the central server selects a set of users and sends them the initial model parameters. Each user provides its local data for training, does the training locally, and sends the local model updates to the central server, while the central server aggregates the final model. Hence, raw data are not shared in this approach, whereas the neural network architecture, parameters, and the intermediate representation of the model (weights, activations, and gradients) are shared between the users and the server during the learning process. With the help of FL, the clients can benefit from obtaining a well-trained ML model without having to send their private data to a central server.

\subsubsection{Definition of FL}
Let's assume their are \textit{N} users $\mathcal{U} = \{u_1, u_2, \ldots, u_n\}$ own their own datasets $ \{\mathcal{D}_1, \mathcal{D}_2, \ldots, \mathcal{D}_n\}$ and each of them cannot directly access to other users' data to expand their own data. %\autoref{fig:federated-learning} shows a basic FL flow that contains these basic steps: 

\begin{inparaenum}[(1)]
    \item The server initializes and sends the initial model to each user
    
    \item Each client trains its own local model $\mathcal{W}_{i}$ on the local data $\mathcal{D}_{i}$  
    
    \item The server collects and aggregates the local models $\mathcal{W}_{i}$ from the clients into an updated global model, and share this global model to all users

    \item Repeat step 2 until the model converges, or the maximum number of the FL rounds has been reached

\end{inparaenum}

\subsection{Homomorphic Encryption}
\label{subsec:he}
\begin{definition}[Homomorphic Encryption]
Let $\mathsf{HE}$ be a (public-key) homomorphic encryption scheme with a quadruple of PPT algorithms $\mathsf{HE = (KeyGen, } \allowbreak \mathsf{Enc, Dec, Eval)}$ such that:
\end{definition}
\begin{itemize}
\item $\mathbf{HE.KeyGen:}$ The key generation algorithm $\left(\mathsf{pk, evk, sk}\right) \leftarrow \mathsf{HE.KeyGen}(1^{\lambda})$ takes as input a unary representation of the security parameter $\lambda$, and outputs  a public key $\mathsf{pk}$, an evaluation key $\mathsf{evk}$ and a private key $\mathsf{sk}$.
\item $\mathbf{HE.Enc:}$ The encryption algorithm ${c \leftarrow \mathsf{HE.Enc}(\mathsf{pk}, x)}$ takes as input the public key $\mathsf{pk}$ and a message $x$ and outputs a ciphertext $c$.
\item $\mathbf{HE.Eval:}$ The algorithm $c_f \leftarrow \mathsf{HE.Eval}({\mathsf{evk}}, f, c_1, \dots, c_n)$ takes as input the evaluation key $\mathsf{evk}$, a function $f$, and a set of $n$ ciphertexts, and outputs a ciphertext $c_f$.
\item $\mathbf{HE.Dec:}$ The decryption algorithm $\mathsf{HE.Dec}({\mathsf{sk}}, c) \rightarrow x$, takes as input the secret key $\mathsf{sk}$ and a ciphertext $c$, and outputs a plaintext $x$.
\end{itemize}
% \vspace{-1.5em}

\subsection{Hybrid Homomorphic Encryption}
\label{subsec: hhe}

\begin{definition}[Hybrid Homomorphic Encryption] Let $\mathsf{HE}$ be a Homomorphic Encryption scheme and $\mathsf{SKE} = (\mathsf{Gen, Enc, Dec})$ be a symmetric-key encryption scheme. Moreover, let $\mathcal{X} = (x_1, \dots, x_n)$ be the message space and $\lambda$ the security parameter. An $\mathsf{HHE}$ scheme consists of five PPT algorithms $\mathsf{HHE = (KeyGen, }$ \allowbreak $\mathsf{Enc, Decomp, Eval, Dec)}$ 
such that:
\end{definition}

\begin{itemize}[left=0pt, nosep]
		\item $\mathbf{HHE.KeyGen}$: The key generation algorithm takes as input a security parameter $\lambda$ and outputs a HE public/private key pair ($\mathsf{pk}$/$\mathsf{sk}$) and a HE evaluation key ($\mathsf{evk}$).

		\item $\mathbf{HHE.Enc}$: This algorithm consists of three steps:
		\begin{itemize}[left=0pt]
			\item $\mathsf{SKE.Gen}$: The SKE generation algorithm takes as input security parameter $\lambda$ and outputs a symmetric key $\mathsf{K}$.
			
                \item $\mathsf{HE.Enc}$: An HE encryption algorithm that takes as input $\mathsf{pk}$ and $\mathsf{K}$, and outputs $c_\mathsf{K}$ -- a homomorphically encrypted representation of the symmetric key $\mathsf{K}$.
			\item $\mathsf{SKE.Enc}$: The SKE encryption algorithm takes as input a message $x$ and $\mathsf{K}$ and outputs a ciphertext $c$.
		\end{itemize}
		\item $\mathbf{HHE.Decomp}$: This algorithm takes as an input the $\mathsf{evk}$, the symmetrically encrypted ciphertext $c$, and the homomorphically encrypted symmetric key $c_\mathsf{K}$, and outputs $c'$ -- a homomorphic encryption of the original message $x$. 

		\item $\mathbf{HHE.Eval}$: It takes as input $n$ homomorphic ciphertexts $c'_n$, where $n \geq 2$, the $\mathsf{evk}$ and a homomorphic function $f$, and outputs a ciphertext $c'_{eval}$ of the evaluation results.

		\item $\mathbf{HHE.Dec}$: This algorithm takes as input a private key $\mathsf{sk}$ and the evaluated ciphertext $c'_{eval}$ and outputs $f(x)$.  

	\end{itemize} 
	%\vspace{-2em}

%The correctness of an HHE scheme follows directly from the correctness of the underlying public-key HE scheme. 

\subsection{Rubato HHE Framework}
\label{subsec: rubato}
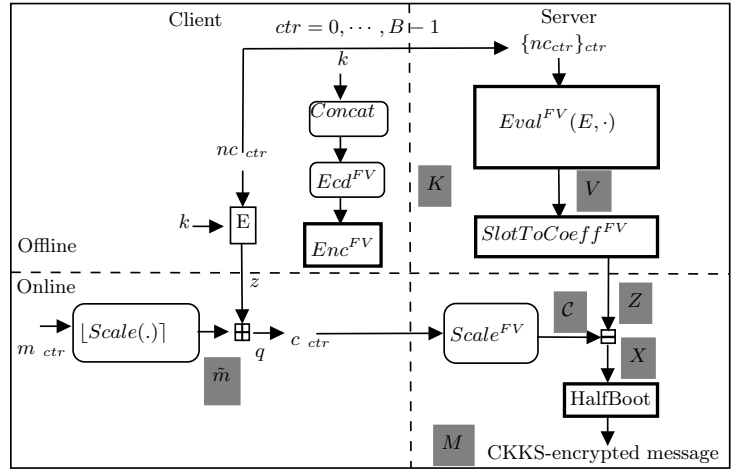
\begin{figure}
 \centering
\vskip 0.2in
\resizebox{0.8\textwidth}{!}
{
\tikzset{every picture/.style={line width=0.75pt}} %set default line width to 0.75pt        

\begin{tikzpicture}[x=0.75pt,y=0.75pt,yscale=-1,xscale=1]
%uncomment if require: \path (0,315); %set diagram left start at 0, and has height of 315

%Shape: Rectangle [id:dp11447973024190661] 
\draw   (221.3,123.75) -- (236.5,123.75) -- (236.5,145.7) -- (221.3,145.7) -- cycle ;
%Rounded Rect [id:dp5755817801259634] 
\draw   (126.5,189) .. controls (126.5,185.13) and (129.63,182) .. (133.5,182) -- (195.4,182) .. controls (199.27,182) and (202.4,185.13) .. (202.4,189) -- (202.4,210) .. controls (202.4,213.87) and (199.27,217) .. (195.4,217) -- (133.5,217) .. controls (129.63,217) and (126.5,213.87) .. (126.5,210) -- cycle ;
%Straight Lines [id:da6221977393678757] 
\draw    (106.5,195.5) -- (122.5,195.5) ;
\draw [shift={(125.5,195.5)}, rotate = 180] [fill={rgb, 255:red, 0; green, 0; blue, 0 }  ][line width=0.08]  [draw opacity=0] (8.93,-4.29) -- (0,0) -- (8.93,4.29) -- cycle    ;
%Straight Lines [id:da22650099495025589] 
\draw  [dash pattern={on 4.5pt off 4.5pt}]  (89,163) -- (522,164) ;
%Straight Lines [id:da9895459082482143] 
\draw    (201.4,198.8) -- (217.4,198.8) ;
\draw [shift={(220.4,198.8)}, rotate = 180] [fill={rgb, 255:red, 0; green, 0; blue, 0 }  ][line width=0.08]  [draw opacity=0] (8.93,-4.29) -- (0,0) -- (8.93,4.29) -- cycle    ;
%Shape: Rectangle [id:dp8064853355445465] 
\draw   (224.2,194.2) -- (228.4,194.2) -- (228.4,199) -- (224.2,199) -- cycle ;
%Shape: Rectangle [id:dp6049796175737072] 
\draw   (228.4,194.2) -- (232.6,194.2) -- (232.6,199) -- (228.4,199) -- cycle ;
%Shape: Rectangle [id:dp3995755741933096] 
\draw   (224.2,199.2) -- (228.4,199.2) -- (228.4,204) -- (224.2,204) -- cycle ;
%Shape: Rectangle [id:dp25844056389281755] 
\draw   (228.4,199.2) -- (232.6,199.2) -- (232.6,204) -- (228.4,204) -- cycle ;

%Straight Lines [id:da4707279762145864] 
\draw    (235,199.6) -- (251,199.6) ;
\draw [shift={(254,199.6)}, rotate = 180] [fill={rgb, 255:red, 0; green, 0; blue, 0 }  ][line width=0.08]  [draw opacity=0] (8.93,-4.29) -- (0,0) -- (8.93,4.29) -- cycle    ;
%Straight Lines [id:da03678359202953008] 
\draw    (273,200) -- (346.5,199.76) ;
\draw [shift={(349.5,199.75)}, rotate = 179.81] [fill={rgb, 255:red, 0; green, 0; blue, 0 }  ][line width=0.08]  [draw opacity=0] (8.93,-4.29) -- (0,0) -- (8.93,4.29) -- cycle    ;
%Rounded Rect [id:dp3901339759634763] 
\draw   (350,189.02) .. controls (350,185.05) and (353.22,181.83) .. (357.18,181.83) -- (399.32,181.83) .. controls (403.28,181.83) and (406.5,185.05) .. (406.5,189.02) -- (406.5,210.57) .. controls (406.5,214.53) and (403.28,217.75) .. (399.32,217.75) -- (357.18,217.75) .. controls (353.22,217.75) and (350,214.53) .. (350,210.57) -- cycle ;
%Shape: Rectangle [id:dp3959398823900009] 
\draw  [line width=1.5]  (266,133.75) -- (312,133.75) -- (312,159.25) -- (266,159.25) -- cycle ;
%Rounded Rect [id:dp4300453590119432] 
\draw   (269.33,101) .. controls (269.33,98.7) and (271.2,96.83) .. (273.5,96.83) -- (309.83,96.83) .. controls (312.13,96.83) and (314,98.7) .. (314,101) -- (314,113.5) .. controls (314,115.8) and (312.13,117.67) .. (309.83,117.67) -- (273.5,117.67) .. controls (271.2,117.67) and (269.33,115.8) .. (269.33,113.5) -- cycle ;
%Rounded Rect [id:dp32531918403024385] 
\draw   (269,64.33) .. controls (269,62.03) and (270.87,60.17) .. (273.17,60.17) -- (311.83,60.17) .. controls (314.13,60.17) and (316,62.03) .. (316,64.33) -- (316,76.83) .. controls (316,79.13) and (314.13,81) .. (311.83,81) -- (273.17,81) .. controls (270.87,81) and (269,79.13) .. (269,76.83) -- cycle ;
%Straight Lines [id:da9188146438540707] 
\draw    (198.4,135.3) -- (214.4,135.3) ;
\draw [shift={(217.4,135.3)}, rotate = 180] [fill={rgb, 255:red, 0; green, 0; blue, 0 }  ][line width=0.08]  [draw opacity=0] (8.93,-4.29) -- (0,0) -- (8.93,4.29) -- cycle    ;
%Straight Lines [id:da9920999360423375] 
\draw    (287.5,118.25) -- (287.5,131.25) ;
\draw [shift={(287.5,134.25)}, rotate = 270] [fill={rgb, 255:red, 0; green, 0; blue, 0 }  ][line width=0.08]  [draw opacity=0] (8.93,-4.29) -- (0,0) -- (8.93,4.29) -- cycle    ;
%Straight Lines [id:da08436901791781937] 
\draw  [dash pattern={on 4.5pt off 4.5pt}]  (329,0.75) -- (330,281) ;
%Shape: Rectangle [id:dp7650403608336916] 
\draw  [line width=1.5]  (368.47,51) -- (479,51) -- (479,100) -- (368.47,100) -- cycle ;
%Shape: Rectangle [id:dp43255773978190115] 
\draw  [line width=1.5]  (368.97,129.58) -- (478.5,129.58) -- (478.5,153.58) -- (368.97,153.58) -- cycle ;
%Shape: Rectangle [id:dp8999461599373285] 
\draw  [line width=1.5]  (422.47,230.25) -- (478.5,230.25) -- (478.5,249.75) -- (422.47,249.75) -- cycle ;
%Shape: Rectangle [id:dp8399969724375513] 
\draw   (444.7,197.75) -- (453,197.75) -- (453,201.75) -- (444.7,201.75) -- cycle ;
%Shape: Rectangle [id:dp49140049738223335] 
\draw   (444.7,202.2) -- (453,202.2) -- (453,205.75) -- (444.7,205.75) -- cycle ;

%Straight Lines [id:da08658188176145298] 
\draw    (448.5,206.75) -- (448.5,219.25) -- (448.5,226.75) ;
\draw [shift={(448.5,229.75)}, rotate = 270] [fill={rgb, 255:red, 0; green, 0; blue, 0 }  ][line width=0.08]  [draw opacity=0] (8.93,-4.29) -- (0,0) -- (8.93,4.29) -- cycle    ;
%Straight Lines [id:da7409453958372497] 
\draw    (406,202) -- (441.7,202.18) ;
\draw [shift={(444.7,202.2)}, rotate = 180.3] [fill={rgb, 255:red, 0; green, 0; blue, 0 }  ][line width=0.08]  [draw opacity=0] (8.93,-4.29) -- (0,0) -- (8.93,4.29) -- cycle    ;
%Straight Lines [id:da5977349557099207] 
\draw    (229,28.25) -- (386,27.76) ;
\draw [shift={(389,27.75)}, rotate = 179.82] [fill={rgb, 255:red, 0; green, 0; blue, 0 }  ][line width=0.08]  [draw opacity=0] (8.93,-4.29) -- (0,0) -- (8.93,4.29) -- cycle    ;
%Straight Lines [id:da962603475151719] 
\draw    (449,154) -- (448.86,195.75) ;
\draw [shift={(448.85,198.75)}, rotate = 270.19] [fill={rgb, 255:red, 0; green, 0; blue, 0 }  ][line width=0.08]  [draw opacity=0] (8.93,-4.29) -- (0,0) -- (8.93,4.29) -- cycle    ;
%Straight Lines [id:da3045695807948594] 
\draw    (228.67,101.67) -- (228.67,120.75) ;
\draw [shift={(228.67,123.75)}, rotate = 270] [fill={rgb, 255:red, 0; green, 0; blue, 0 }  ][line width=0.08]  [draw opacity=0] (8.93,-4.29) -- (0,0) -- (8.93,4.29) -- cycle    ;
%Straight Lines [id:da7165293140780585] 
\draw    (287.5,81.25) -- (287.5,94.25) ;
\draw [shift={(287.5,97.25)}, rotate = 270] [fill={rgb, 255:red, 0; green, 0; blue, 0 }  ][line width=0.08]  [draw opacity=0] (8.93,-4.29) -- (0,0) -- (8.93,4.29) -- cycle    ;
%Straight Lines [id:da04495828920351763] 
\draw    (287.5,43.25) -- (287.5,56.25) ;
\draw [shift={(287.5,59.25)}, rotate = 270] [fill={rgb, 255:red, 0; green, 0; blue, 0 }  ][line width=0.08]  [draw opacity=0] (8.93,-4.29) -- (0,0) -- (8.93,4.29) -- cycle    ;
%Straight Lines [id:da12958989220999606] 
\draw    (419,33.75) -- (419.28,49) ;
\draw [shift={(419.33,52)}, rotate = 268.95] [fill={rgb, 255:red, 0; green, 0; blue, 0 }  ][line width=0.08]  [draw opacity=0] (8.93,-4.29) -- (0,0) -- (8.93,4.29) -- cycle    ;
%Straight Lines [id:da16642631165497046] 
\draw    (419,100) -- (419.3,127.67) ;
\draw [shift={(419.33,130.67)}, rotate = 269.38] [fill={rgb, 255:red, 0; green, 0; blue, 0 }  ][line width=0.08]  [draw opacity=0] (8.93,-4.29) -- (0,0) -- (8.93,4.29) -- cycle    ;
%Straight Lines [id:da15115560767375524] 
\draw    (448.5,249.75) -- (448.5,257.25) -- (448.5,264.75) ;
\draw [shift={(448.5,267.75)}, rotate = 270] [fill={rgb, 255:red, 0; green, 0; blue, 0 }  ][line width=0.08]  [draw opacity=0] (8.93,-4.29) -- (0,0) -- (8.93,4.29) -- cycle    ;
%Straight Lines [id:da7203556680333805] 
\draw    (229,28.25) -- (228.67,91) ;
%Straight Lines [id:da6848531671114427] 
\draw    (228,146) -- (228.38,191.2) ;
\draw [shift={(228.4,194.2)}, rotate = 269.52] [fill={rgb, 255:red, 0; green, 0; blue, 0 }  ][line width=0.08]  [draw opacity=0] (8.93,-4.29) -- (0,0) -- (8.93,4.29) -- cycle    ;
%Shape: Rectangle [id:dp4743044931517263] 
\draw   (89,1) -- (522,1) -- (522,281) -- (89,281) -- cycle ;

% Text Node
\draw (91.4,205.5) node [anchor=north west][inner sep=0.75pt]    {$m_{\ }{}_{ctr}$};
% Text Node
\draw (128.5,192.4) node [anchor=north west][inner sep=0.75pt]    {$\lfloor Scale( .) \rceil $};
% Text Node
\draw (256.18,200.2) node [anchor=north west][inner sep=0.75pt]    {$c_{\ }{}_{ctr}$};
% Text Node
\draw (352,192.42) node [anchor=north west][inner sep=0.75pt]    {$Scale^{FV}$};
% Text Node
\draw (91.5,141) node [anchor=north west][inner sep=0.75pt]   [align=left] {Offline};
% Text Node
\draw (91,166) node [anchor=north west][inner sep=0.75pt]   [align=left] {Online};
% Text Node
\draw (223.3,126.75) node [anchor=north west][inner sep=0.75pt]   [align=left] {E};
% Text Node
\draw (268.2,141.03) node [anchor=north west][inner sep=0.75pt]    {$Enc^{FV}$};
% Text Node
\draw (271.2,100.03) node [anchor=north west][inner sep=0.75pt]    {$Ecd^{FV}$};
% Text Node
\draw (267.7,59.53) node [anchor=north west][inner sep=0.75pt]    {$Concat$};
% Text Node
\draw (424.47,233.25) node [anchor=north west][inner sep=0.75pt]   [align=left] {{\footnotesize HalfBoot}};
% Text Node
\draw (381.2,63.03) node [anchor=north west][inner sep=0.75pt]    {${\displaystyle Eval^{FV}( E,\cdot )}$};
% Text Node
\draw (370.7,131.7) node [anchor=north west][inner sep=0.75pt]    {$SlotToCoeff^{FV}$};
% Text Node
\draw (394.2,18.03) node [anchor=north west][inner sep=0.75pt]    {$\{nc_{ctr}\}_{ctr}$};
% Text Node
\draw (182.5,4.5) node [anchor=north west][inner sep=0.75pt]   [align=left] {Client};
% Text Node
\draw (407,3) node [anchor=north west][inner sep=0.75pt]   [align=left] {Server};
% Text Node
\draw (247,8.92) node [anchor=north west][inner sep=0.75pt]    {$ctr=0,\cdots ,B-1$};
% Text Node
\draw (284.2,28.03) node [anchor=north west][inner sep=0.75pt]    {$k$};
% Text Node
\draw (188.7,126.53) node [anchor=north west][inner sep=0.75pt]    {$k$};
% Text Node
\draw (374.47,265.58) node [anchor=north west][inner sep=0.75pt]   [align=left] {{\footnotesize CKKS-encrypted message}};
% Text Node
\draw (211.33,85.5) node [anchor=north west][inner sep=0.75pt]    {$nc_{\ }{}_{ctr}$};
% Text Node
\draw  [color={rgb, 255:red, 128; green, 128; blue, 128 }  ,draw opacity=1 ][fill={rgb, 255:red, 128; green, 128; blue, 128 }  ,fill opacity=1 ]  (416.53,175) -- (435.53,175) -- (435.53,199) -- (416.53,199) -- cycle  ;
\draw (419.53,179.4) node [anchor=north west][inner sep=0.75pt]    {$\mathcal{C}$};
% Text Node
\draw  [color={rgb, 255:red, 128; green, 128; blue, 128 }  ,draw opacity=1 ][fill={rgb, 255:red, 128; green, 128; blue, 128 }  ,fill opacity=1 ]  (334.87,99) -- (353.87,99) -- (353.87,123) -- (334.87,123) -- cycle  ;
\draw (337.87,103.4) node [anchor=north west][inner sep=0.75pt]    {$K$};
% Text Node
\draw  [color={rgb, 255:red, 128; green, 128; blue, 128 }  ,draw opacity=1 ][fill={rgb, 255:red, 128; green, 128; blue, 128 }  ,fill opacity=1 ]  (455.87,169.67) -- (474.87,169.67) -- (474.87,193.67) -- (455.87,193.67) -- cycle  ;
\draw (458.87,174.07) node [anchor=north west][inner sep=0.75pt]    {$Z$};
% Text Node
\draw  [color={rgb, 255:red, 128; green, 128; blue, 128 }  ,draw opacity=1 ][fill={rgb, 255:red, 128; green, 128; blue, 128 }  ,fill opacity=1 ]  (457,202.75) -- (477,202.75) -- (477,226.75) -- (457,226.75) -- cycle  ;
\draw (460,207.15) node [anchor=north west][inner sep=0.75pt]    {$X$};
% Text Node
\draw  [color={rgb, 255:red, 128; green, 128; blue, 128 }  ,draw opacity=1 ][fill={rgb, 255:red, 128; green, 128; blue, 128 }  ,fill opacity=1 ]  (343.87,255.33) -- (367.87,255.33) -- (367.87,279.33) -- (343.87,279.33) -- cycle  ;
\draw (346.87,259.73) node [anchor=north west][inner sep=0.75pt]    {$M$};
% Text Node
\draw  [color={rgb, 255:red, 128; green, 128; blue, 128 }  ,draw opacity=1 ][fill={rgb, 255:red, 128; green, 128; blue, 128 }  ,fill opacity=1 ]  (430.33,102) -- (450.33,102) -- (450.33,126) -- (430.33,126) -- cycle  ;
\draw (433.33,106.4) node [anchor=north west][inner sep=0.75pt]    {$V$};
% Text Node
\draw  [color={rgb, 255:red, 128; green, 128; blue, 128 }  ,draw opacity=1 ][fill={rgb, 255:red, 128; green, 128; blue, 128 }  ,fill opacity=1 ]  (205.87,214.67) -- (226.87,214.67) -- (226.87,242.67) -- (205.87,242.67) -- cycle  ;
\draw (208.87,219.07) node [anchor=north west][inner sep=0.75pt]    {$\tilde{m}$};
% Text Node
\draw (234.6,207.4) node [anchor=north west][inner sep=0.75pt]    {$q$};
% Text Node
\draw (231.27,165.27) node [anchor=north west][inner sep=0.75pt]    {$z$};
\end{tikzpicture}}
    \caption{The RtF transciphering framework.}
    \label{fig:enter-label}
    \vskip -0.2in
\end{figure}

We present the Rubato and RtF HHE framework~\cite{ha2022rubato} in~\autoref{fig:enter-label}. HE evaluation are performed in the boxes with thick lines. Operations in the boxes with rounded corners do not use any secret information.

\textbf{Initialization} The RtF transciphering framework uses \textit{both FV and CKKS schemes}. For a fixed security parameter $\lambda$, set parameters such as the degree of the polynomial modulus and the ciphertext moduli, and \textit{generate the public-private key pair} to satisfy the desired security level $\lambda$. For a \textit{secret key $k \in \mathbb{Z}^n_q$ used for a symmetric cipher $\mathbf{E}$}, the client computes the \textit{FV-ciphertext $\mathcal{K}$ of the symmetric secret key $k$} and sends it to the server.

\textbf{Client-side Computation}:
\begin{itemize}[left=0pt, nosep]
    \item \textbf{Offline Phase:} The client \textit{generates a keystream $\mathbf{z} \in \mathbb{Z}_q^n$ from the cipher $\mathsf{E}$}, taking the nonce $\mathsf{nc} \in \{0,1\}^\lambda$ and the secret key $\mathbf{k} \in \mathbb{Z}_q^n$ as inputs.
    \item \textbf{Online Phase:} 
    First, the client scales up the plaintext message $m$ (in floating point) by a scaling factor $\Delta$ to obtain a quantized message:
    $\widetilde{\mathbf{m}} = \lfloor \Delta \cdot \mathbf{m} \rceil \in \mathbb{Z}_q^n.$
    
    Then, the client performs addition modulo $q$ between the keystream $\mathbf{z}$ and the scaled message $\widetilde{\mathbf{m}}$, resulting in a symmetric ciphertext:
    $\mathbf{c} = [\widetilde{\mathbf{m}} + \mathbf{z}]_q.$
    The client sends $\mathbf{c}$ along with the nonce $\mathsf{nc}$ to the server.
\end{itemize}

\textbf{Server-side Computation}:
\begin{itemize}[left=0pt, nosep]
    \item \textbf{Offline Phase:}
    \begin{itemize}[left=0pt]
        \item First, the server \textit{evaluates the keystream using a tuple of nonces $(\mathsf{nc}_0,\ldots,\mathsf{nc}_{B-1})$ and the FV-encrypted symmetric key $\mathcal{K}$}. This results in an FV-ciphertext $\mathcal{V}$ containing the keystreams of $\mathsf{E}$ in its slots.
        \item Then, the server applies a linear transformation $\mathsf{SlotToCoeff}^{\mathsf{FV}}$ to move the data from the slots to the coefficients, obtaining an FV-ciphertext $\mathcal{Z}$ containing the keystreams in its coefficients. \textit{All of the homomorphic evaluations above are done in the FV scheme.}
    \end{itemize}
    
    \item \textbf{Online Phase:}
    \begin{itemize}[left=0pt]
        \item During online phase, the server first \textit{scales up the symmetric ciphertext $\mathbf{c}$ into the FV ciphertext space}, resulting in an FV-ciphertext $\mathcal{C}$ containing the symmetric ciphertexts in its coefficients.
        \item Then, it subtracts homomorphically the evaluated keystream $\mathcal{Z}$ from $\mathcal{C}$:
        $\mathcal{X} = \mathcal{C} - \mathcal{Z},$
        yielding an FV-ciphertext $\mathcal{X}$ having scaled messages in its coefficients.
        \item The final step is $\mathsf{HalfBoot}$, a modified bootstrapping process in the $\mathsf{RtF}$ framework. Given $\mathcal{X}$ as input, it outputs a CKKS-ciphertext:
        $\mathsf{HalfBoot}(\mathcal{X}) = \mathcal{M},$
        where $\mathcal{M}$ contains the CKKS-encrypted messages in its slots.
    \end{itemize}
\end{itemize}

\section{\flhhe{}}
\label{sec: FLHHE}

\subsection{System Model}
\label{subsec:system-model}
In this section, we introduce our system model by explicitly describing our protocol's %the%
main entities and 
their capabilities. 
\begin{itemize}[left=0pt, nosep]
    \item \textbf{Trusted key dealer}: The trusted key dealer creates and distributes all the HHE keys to the clients along with an HE encryption $c_\mathsf{K_i}$ of the symmetric key to the $\mathbf{CSP}$.
        
    \item \textbf{Client}: Let $\mathcal{C} = \{c_1, \ldots, C_n\}$ be the set of all clients. Each client uses the unique symmetric key $\mathsf{K_i}$ locally and encrypts their data. The generated ciphertexts are then outsourced to the $\mathbf{CSP}$.

    \item \textbf{Cloud Service Provider (CSP)}: %An entity that is p
    Primarily responsible for gathering symmetrically encrypted data from multiple users. The $\mathbf{CSP}$ is tasked with converting the symmetrically encrypted data into homomorphic ciphertexts and, upon request, performing blind operations on them. 
\end{itemize}

\subsection{Threat Model}
\label{subsec:threat-model}

We consider a semi-honest (honest-but-curious) threat model, where both clients and server, follow the protocol correctly but may attempt to infer private information from the data they receive. To ensure privacy in this setting, the system must satisfy the following requirements:

\begin{inparaenum}
    [\it (i)] \item Client should only access their own local models and the final aggregated model, without visibility into other clients' models or data \item The server must receive encrypted models and secure parameters, without learning the raw data. \item All client-side data remains strictly confidential and protected from both the server and other clients.
\end{inparaenum}

\section{Methodology}
\label{sec:methodology}

\begin{figure}
\centering
\resizebox{0.8\textwidth}{!}
{
\tikzset{every picture/.style={line width=0.75pt}}

\begin{tikzpicture}[x=0.75pt,y=0.75pt,yscale=-1,xscale=1]
\draw  [color={rgb, 255:red, 245; green, 166; blue, 35 }  ,draw opacity=1 ] (49,72.25) -- (49,78.75) .. controls (49,79.72) and (44.08,80.5) .. (38,80.5) .. controls (31.92,80.5) and (27,79.72) .. (27,78.75) -- (27,72.25)(49,72.25) .. controls (49,73.22) and (44.08,74) .. (38,74) .. controls (31.92,74) and (27,73.22) .. (27,72.25) .. controls (27,71.28) and (31.92,70.5) .. (38,70.5) .. controls (44.08,70.5) and (49,71.28) .. (49,72.25) -- cycle ;
%Straight Lines [id:da6893569676936696] 
\draw [color={rgb, 255:red, 245; green, 166; blue, 35 }  ,draw opacity=1 ]   (53.89,74.84) -- (68.66,74.96) ;
\draw [shift={(71.66,74.99)}, rotate = 180.49] [fill={rgb, 255:red, 245; green, 166; blue, 35 }  ,fill opacity=1 ][line width=0.08]  [draw opacity=0] (5.36,-2.57) -- (0,0) -- (5.36,2.57) -- cycle    ;
%Flowchart: Magnetic Disk [id:dp8461810725101482] 
\draw  [color={rgb, 255:red, 139; green, 87; blue, 42 }  ,draw opacity=1 ] (48.61,140.98) -- (48.61,147.16) .. controls (48.61,148.08) and (43.52,148.83) .. (37.25,148.83) .. controls (30.97,148.83) and (25.88,148.08) .. (25.88,147.16) -- (25.88,140.98)(48.61,140.98) .. controls (48.61,141.9) and (43.52,142.65) .. (37.25,142.65) .. controls (30.97,142.65) and (25.88,141.9) .. (25.88,140.98) .. controls (25.88,140.06) and (30.97,139.32) .. (37.25,139.32) .. controls (43.52,139.32) and (48.61,140.06) .. (48.61,140.98) -- cycle ;
%Straight Lines [id:da9397073623663691] 
\draw [color={rgb, 255:red, 139; green, 87; blue, 42 }  ,draw opacity=1 ]   (51.61,145.16) -- (66.38,145.29) ;
\draw [shift={(69.38,145.32)}, rotate = 180.49] [fill={rgb, 255:red, 139; green, 87; blue, 42 }  ,fill opacity=1 ][line width=0.08]  [draw opacity=0] (5.36,-2.57) -- (0,0) -- (5.36,2.57) -- cycle    ;
%Flowchart: Magnetic Disk [id:dp5173196224270713] 
\draw  [color={rgb, 255:red, 128; green, 128; blue, 128 }  ,draw opacity=1 ] (48,205.33) -- (48,212.11) .. controls (48,213.12) and (43.47,213.94) .. (37.89,213.94) .. controls (32.3,213.94) and (27.78,213.12) .. (27.78,212.11) -- (27.78,205.33)(48,205.33) .. controls (48,206.34) and (43.47,207.15) .. (37.89,207.15) .. controls (32.3,207.15) and (27.78,206.34) .. (27.78,205.33) .. controls (27.78,204.32) and (32.3,203.5) .. (37.89,203.5) .. controls (43.47,203.5) and (48,204.32) .. (48,205.33) -- cycle ;
%Straight Lines [id:da943865509406428] 
\draw [color={rgb, 255:red, 128; green, 128; blue, 128 }  ,draw opacity=1 ]   (52.01,209.2) -- (66.78,209.33) ;
\draw [shift={(69.78,209.35)}, rotate = 180.49] [fill={rgb, 255:red, 128; green, 128; blue, 128 }  ,fill opacity=1 ][line width=0.08]  [draw opacity=0] (5.36,-2.57) -- (0,0) -- (5.36,2.57) -- cycle    ;
%Image [id:dp08981853792402206] 
\draw (95.49,212.54) node  {\includegraphics[width=33.76pt,height=26.18pt]{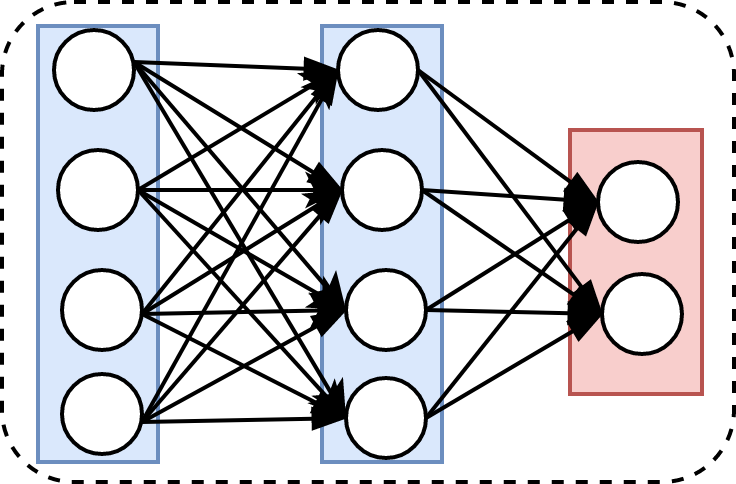}};
%Straight Lines [id:da18858540159390602] 
\draw  [dash pattern={on 4.5pt off 4.5pt}]  (170.5,52) -- (170.2,279.8) ;
%Flowchart: Decision [id:dp7935457103943238] 
\draw   (142,60.5) -- (159,77.5) -- (142,94.5) -- (125,77.5) -- cycle ;
%Flowchart: Decision [id:dp8082189250637501] 
\draw   (142,123.5) -- (159,140.5) -- (142,157.5) -- (125,140.5) -- cycle ;
%Flowchart: Decision [id:dp3545256269081176] 
\draw   (143.33,195.83) -- (160.33,212.83) -- (143.33,229.83) -- (126.33,212.83) -- cycle ;
%Straight Lines [id:da37173420863616513] 
\draw    (260.33,0.67) -- (260.2,277.4) ;
%Flowchart: Decision [id:dp8589385047065475] 
\draw   (213.25,57.25) -- (235,76.88) -- (213.25,96.5) -- (191.5,76.88) -- cycle ;
%Flowchart: Decision [id:dp5168653445135287] 
\draw   (215.75,193.25) -- (237.5,212.88) -- (215.75,232.5) -- (194,212.88) -- cycle ;
%Flowchart: Decision [id:dp9883815024075685] 
\draw   (217.25,119.75) -- (239,139.38) -- (217.25,159) -- (195.5,139.38) -- cycle ;
%Flowchart: Decision [id:dp2809022995914481] 
\draw   (302.25,78.25) -- (324,97.88) -- (302.25,117.5) -- (280.5,97.88) -- cycle ;
%Flowchart: Decision [id:dp678793552034723] 
\draw   (346.75,78.25) -- (368.5,97.88) -- (346.75,117.5) -- (325,97.88) -- cycle ;
%Flowchart: Decision [id:dp9866279726412692] 
\draw   (391.25,78.25) -- (413,97.88) -- (391.25,117.5) -- (369.5,97.88) -- cycle ;
%Flowchart: Decision [id:dp03228262359021361] 
\draw   (302.75,161) -- (324.5,180.63) -- (302.75,200.25) -- (281,180.63) -- cycle ;
%Flowchart: Decision [id:dp5102871576717191] 
\draw   (347.75,160.75) -- (369.5,180.38) -- (347.75,200) -- (326,180.38) -- cycle ;
%Flowchart: Decision [id:dp8237391885510728] 
\draw   (391.25,160.75) -- (413,180.38) -- (391.25,200) -- (369.5,180.38) -- cycle ;
%Curve Lines [id:da2929968887204015] 
\draw    (235,76.88) .. controls (284.74,74.41) and (228.73,95.24) .. (278.17,97.77) ;
\draw [shift={(280.5,97.88)}, rotate = 182.14] [fill={rgb, 255:red, 0; green, 0; blue, 0 }  ][line width=0.08]  [draw opacity=0] (8.93,-4.29) -- (0,0) -- (8.93,4.29) -- cycle    ;
%Curve Lines [id:da32984509814199736] 
\draw    (239,139.38) .. controls (288.74,136.91) and (228.85,97.1) .. (278.17,97.81) ;
\draw [shift={(280.5,97.88)}, rotate = 182.14] [fill={rgb, 255:red, 0; green, 0; blue, 0 }  ][line width=0.08]  [draw opacity=0] (8.93,-4.29) -- (0,0) -- (8.93,4.29) -- cycle    ;
%Curve Lines [id:da6370490893247447] 
\draw    (237.5,212.88) .. controls (287.24,210.41) and (228.8,99.28) .. (278.17,97.86) ;
\draw [shift={(280.5,97.88)}, rotate = 182.14] [fill={rgb, 255:red, 0; green, 0; blue, 0 }  ][line width=0.08]  [draw opacity=0] (8.93,-4.29) -- (0,0) -- (8.93,4.29) -- cycle    ;
%Straight Lines [id:da25199911613117865] 
\draw    (302.5,141.25) -- (302.71,158) ;
\draw [shift={(302.75,161)}, rotate = 269.27] [fill={rgb, 255:red, 0; green, 0; blue, 0 }  ][line width=0.08]  [draw opacity=0] (8.93,-4.29) -- (0,0) -- (8.93,4.29) -- cycle    ;
%Rounded Rect [id:dp7311350194268658] 
\draw   (301,233.2) .. controls (301,228.81) and (304.56,225.25) .. (308.95,225.25) -- (391.05,225.25) .. controls (395.44,225.25) and (399,228.81) .. (399,233.2) -- (399,257.05) .. controls (399,261.44) and (395.44,265) .. (391.05,265) -- (308.95,265) .. controls (304.56,265) and (301,261.44) .. (301,257.05) -- cycle ;
%Curve Lines [id:da5406722293518883] 
\draw    (302.25,200) .. controls (345.69,199.21) and (346.75,212.34) .. (346.55,221.88) ;
\draw [shift={(346.5,224.75)}, rotate = 270] [fill={rgb, 255:red, 0; green, 0; blue, 0 }  ][line width=0.08]  [draw opacity=0] (8.93,-4.29) -- (0,0) -- (8.93,4.29) -- cycle    ;
%Curve Lines [id:da9620805804537033] 
\draw    (391.25,200) .. controls (353.23,201.58) and (347.4,212.67) .. (346.61,221.8) ;
\draw [shift={(346.5,224.75)}, rotate = 270] [fill={rgb, 255:red, 0; green, 0; blue, 0 }  ][line width=0.08]  [draw opacity=0] (8.93,-4.29) -- (0,0) -- (8.93,4.29) -- cycle    ;
%Straight Lines [id:da27504089296425793] 
\draw    (347.75,200) -- (346.65,221.75) ;
\draw [shift={(346.5,224.75)}, rotate = 272.89] [fill={rgb, 255:red, 0; green, 0; blue, 0 }  ][line width=0.08]  [draw opacity=0] (8.93,-4.29) -- (0,0) -- (8.93,4.29) -- cycle    ;
%Straight Lines [id:da29325733866281645] 
\draw    (10.4,280.8) -- (319.8,279.8) ;
%Straight Lines [id:da9736196710887877] 
\draw    (11.33,340.17) -- (320.67,339.83) ;
%Straight Lines [id:da6862591015855223] 
\draw    (11.33,340.17) -- (10.4,280.8) ;
%Straight Lines [id:da9162985555072887] 
\draw    (320.67,339.83) -- (319.8,279.8) ;
%Straight Lines [id:da8160569254923109] 
\draw    (302.25,117.5) -- (302.33,128.83) ;
%Straight Lines [id:da8098919486288936] 
\draw    (347.83,140.58) -- (348.05,157.33) ;
\draw [shift={(348.08,160.33)}, rotate = 269.27] [fill={rgb, 255:red, 0; green, 0; blue, 0 }  ][line width=0.08]  [draw opacity=0] (8.93,-4.29) -- (0,0) -- (8.93,4.29) -- cycle    ;
%Straight Lines [id:da3636753110838833] 
\draw    (347.58,116.83) -- (347.67,130.5) ;
%Straight Lines [id:da567652919576391] 
\draw    (391.17,141.92) -- (391.38,158.67) ;
\draw [shift={(391.42,161.67)}, rotate = 269.27] [fill={rgb, 255:red, 0; green, 0; blue, 0 }  ][line width=0.08]  [draw opacity=0] (8.93,-4.29) -- (0,0) -- (8.93,4.29) -- cycle    ;
%Straight Lines [id:da05555402337464044] 
\draw    (390.92,118.17) -- (391,131.83) ;
%Image [id:dp5244830977596842] 
\draw (96.49,142.54) node  {\includegraphics[width=33.76pt,height=26.18pt]{images/ml.png}};
%Image [id:dp6316771921965076] 
\draw (98.49,74.54) node  {\includegraphics[width=33.76pt,height=26.18pt]{images/ml.png}};

% Text Node
\draw (10,55.82) node [anchor=north west][inner sep=0.75pt]    {$Client\ 1$};
% Text Node
\draw (10,123.82) node [anchor=north west][inner sep=0.75pt]    {$Client\ 2$};
% Text Node
\draw (10,186.82) node [anchor=north west][inner sep=0.75pt]    {$Client\ 3$};
% Text Node
\draw (314.83,4.98) node [anchor=north west][inner sep=0.75pt]    {$Server$};
% Text Node
\draw (41.58,104.13) node   [align=left] {\begin{minipage}[lt]{57.69pt}\setlength\topsep{0pt}
{\footnotesize  \ \ \ \ \ MNIST}\\{\footnotesize excluded 1,2,3}
\end{minipage}};
% Text Node
\draw (42.58,165.13) node   [align=left] {\begin{minipage}[lt]{57.69pt}\setlength\topsep{0pt}
{\footnotesize  \ \ \ \ \ MNIST}\\{\footnotesize excluded 4,5,6}
\end{minipage}};
% Text Node
\draw (40.58,237.13) node   [align=left] {\begin{minipage}[lt]{57.69pt}\setlength\topsep{0pt}
{\footnotesize  \ \ \ \ \ MNIST}\\{\footnotesize excluded 7,8,9}
\end{minipage}};
% Text Node
\draw (131,67) node [anchor=north west][inner sep=0.75pt]    {$A1$};
% Text Node
\draw (131,130) node [anchor=north west][inner sep=0.75pt]    {$A2$};
% Text Node
\draw (131.33,202.33) node [anchor=north west][inner sep=0.75pt]    {$An$};
% Text Node
\draw (85.92,39.04) node   [align=left] {\begin{minipage}[lt]{47.26pt}\setlength\topsep{0pt}
Plaintext
\end{minipage}};
% Text Node
\draw (213.42,43.71) node   [align=left] {\begin{minipage}[lt]{47.26pt}\setlength\topsep{0pt}
Symmetric Encryption
\end{minipage}};
% Text Node
\draw (200,67) node [anchor=north west][inner sep=0.75pt]    {$\widehat{A1}$};
% Text Node
\draw (202.5,202) node [anchor=north west][inner sep=0.75pt]    {$\widehat{An}$};
% Text Node
\draw (204,130.5) node [anchor=north west][inner sep=0.75pt]    {$\widehat{A2}$};
% Text Node
\draw (292,87) node [anchor=north west][inner sep=0.75pt]    {$\widehat{A1}$};
% Text Node
\draw (336.5,87) node [anchor=north west][inner sep=0.75pt]    {$\widehat{A2}$};
% Text Node
\draw (382,87) node [anchor=north west][inner sep=0.75pt]    {$\widehat{An}$};
% Text Node
\draw (292,169.5) node [anchor=north west][inner sep=0.75pt]    {$\overline{A1}$};
% Text Node
\draw (336,169.5) node [anchor=north west][inner sep=0.75pt]    {$\overline{A2}$};
% Text Node
\draw (382,169.5) node [anchor=north west][inner sep=0.75pt]    {$\overline{An}$};
% Text Node
\draw (305,128.58) node [anchor=north west][inner sep=0.75pt]   [align=left] {Transciphering};
% Text Node
\draw (354,245.88) node   [align=left] {\begin{minipage}[lt]{68pt}\setlength\topsep{0pt}
  HE Aggregation
\end{minipage}};
% Text Node
\draw (19.26,285) node [anchor=north west][inner sep=0.75pt]    {$Ai$};
% Text Node
\draw (19.26,300) node [anchor=north west][inner sep=0.75pt]    {$\widehat{Ai}$};
% Text Node
\draw (18.26,318.3) node [anchor=north west][inner sep=0.75pt]    {$\overline{Ai}$};
% Text Node
\draw (44.14,300.62) node [anchor=north west][inner sep=0.75pt]   [align=left] {Symmetric encrypted model};
% Text Node
\draw (43.85,318.62) node [anchor=north west][inner sep=0.75pt]   [align=left] {Homomorphic encrypted model};
% Text Node
\draw (44.64,285.62) node [anchor=north west][inner sep=0.75pt]   [align=left] {Plaintext model model};
% Text Node
\draw (114.83,1.33) node [anchor=north west][inner sep=0.75pt]    {$Client$};
% Text Node
\draw (337.92,45.71) node   [align=left] {\begin{minipage}[lt]{47.26pt}\setlength\topsep{0pt}
Ciphertext
\end{minipage}};
\end{tikzpicture}}
\caption{FL Framework with HHE for Secure Model Aggregation}
\label{fig:protocol}
\end{figure}

Our proposed FL system is mainly composed of two key entities: a central model aggregation server and multiple distributed clients, as shown in~\autoref{fig:protocol}. The server is responsible for securely aggregating encrypted local models submitted by the clients, while each client trains a local model on its private dataset and transmits the encrypted model along with secure parameters to the server. Through the collaborative process, an optimal global model is constructed iteratively over multiple training rounds. Unlike conventional FL frameworks, our approach uses HHE to improve privacy and reduce communication overhead between clients and server. The specific construction of the proposed protocol is explained in detail in the following:

\textbf{Initialization}: Before training starts, the clients receive the symmetric key, HE private / public keys, and the server receive HE ciphertext of the symmetric key and HE evaluation key from the key dealer. The server selects participating clients, defines the model architecture and key parameters like learning rate, batch size, number of local epochs. All parties establish secure communication channels and synchronize these parameters. Each client then receives an initial plaintext model from the server in preparation for local training.

\textbf{Local Training}: Each client trains local model on private data, resulting in an updated set of model parameters in plaintext. These models reflect client-specific learning based on their local data. After training, clients encrypt their local model using symmetric encryption. The symmetric encrypted models are then sent to the server.

\textbf{Server Aggregation}: Upon receiving encrypted models, the server uses HHE's transcipher and evaluation primitives to homomorphically convert and aggregate the model. The FL Server aggregates using the most simple weight averaging algorithm \simpavg{} in the HE domain, where each client's model contributes equally, regardless of its dataset size:
%\begin{equation}
    $w_{\text {global }}=\frac{1}{K} \sum_{k=1}^{K} w_{k}.$
%\end{equation}
Here, $K$ is the number of clients, and $w_k$ is the model weights from client $k$. We use simple FL averaging algorithm since our goal is to demonstrate that under HHE, accuracy of the aggregated model is comparable with one produced under plaintext, not to demonstrate how accurate our FL model is.

\section{Experiment}
\label{sec:experiment}

In this section, we evaluate the performance of the \flhhe{} protocol, focusing on its computational and communication cost, and compare its performance against the same FL procedure using a plain CKKS HE scheme. 
Below is the detailed configurations of our experimental environment:
\begin{itemize}[left=0pt, nosep]
        \item \textbf{Hardware}: Our experimental testbed is a commercial Macbook Pro laptop with an Apple M4 Max with 16 CPU cores, integrated GPU of 40 cores, and 128 GB of RAM. All trainings and evaluation are done in a simulated environment under this testbed.
        \item \textbf{Software}: We use the pytorch library %\footnote{\url{http://pypi.org/project/torch/}} 
        to train the neural networks locally for the users, the Lattigo v6 %\footnote{\url{https://pkg.go.dev/github.com/tuneinsight/lattigo/v6}}  
        for plain CKKS, and the Rubato %\footnote{\url{https://github.com/KAIST-CryptLab/Rubato}} 
        combined with the RtF transciphering framework %\footnote{\url{https://github.com/KAIST-CryptLab/RtF-Transciphering}} 
        for our HHE implementations
        \item \textbf{Dataset}: We used the classic MNIST dataset of 28x28 handwritten images to evaluate the performance of our method. We process the raw MNIST dataset, which contains 60,000 samples in total, into 3 partitions for training, each holds by a client:
        \begin{itemize}[left=0pt]
            \item Partition 1: Exclude labels 1, 3, 7 (40,862 samples)
            %Does not contain examples with labels 1, 3, 7 (40,862 samples)
            \item Partition 2: Exclude labels 2, 5, 8  (42,770 samples)
            %Does not contain examples with labels 2, 5, 8  (42,770 samples)
            \item Partition 3: Exclude labels 4, 6, 9 (42,291 samples)
            %Does not contain examples with labels 4, 6, 9 (42,291 samples)
        \end{itemize}
        \item \textbf{Neural Network}: We use a fully connected neural network that consists of two linear layers (without bias terms) and a ReLU activation function in between. The network takes a batch of flattened MNIST images, each with length 784, processes it through a hidden layer of 32 neurons, and produces 10 output values (one for each digit 0-9). The plaintext weights are then flattened into 1D vectors before being encrypted in HE and HHE domains. The values in these weight vectors are floating point in the range of $[-1, 1]$. The weight vectors are in the JSON format for plaintext version, or in binaries for the encrypted versions, for data persistence and communication.
\end{itemize}

% \subsection{Computation Analysis}
\textbf{Computation Analysis}: We report each party's \flhhe{} performance averaged over 10 runs (see~\autoref{table:computation-time}).
% RUBATO128L
% ->> blockSize = 64
% ->> outputSize = 2
% ->> numRound = 2
% ->> plainModulus = 33292289
% ->> sigma = 1.635663
% ->> params.N() = 65536  // ring degree
% ->> params.Slots() = 32768  // number of available plaintext slots
% \vskip -0.15in
\begin{table}[ht]
\centering
\caption{Computation Time Breakdown (in seconds)}
\label{table:computation-time}
\resizebox{0.7\textwidth}{!}
{
\begin{tabular}{@{}l l r@{}}
\toprule
\textbf{Role} & \textbf{Task} & \textbf{Time (sec)} \\
\midrule
\multirow{2}{*}{Key Dealer} 
    & Key generation (once) & 57.45 \\
    & Load keys & 20.53 \\
    & Symmetric key encryption & 5.33 \\
\midrule
\multirow{4}{*}{Client} 
    & Decrypt global model & 0.07 \\
    & Local training & 2.73 \\
    & Keystream generation (offline) & 0.55 \\
    & Symmetric encryption (online) & 0.008 \\
    & \textit{Total per client} & \textit{3.32} \\
\midrule
\multirow{3}{*}{Server (per client)} 
    & Produce $\mathcal{Z}$ (offline) & 32.94 \\
    & Transcipher $\mathcal{M}$ (online) & 6.87 \\
    & HE aggregation (online) & 0.003 \\
\bottomrule
\end{tabular}}
\end{table}

Hence, server processing time per client is 39.81~sec.
% for each client, it takes the server 39.81 sec to process.
With these results, we can see that the initialization/setup cost is quite heavy but is front-loaded, only needed to do once, and scales well with more rounds and clients. For clients, HHE introduces almost zero latency overhead at the client-side since symmetric encryption during the online phase only takes 0.008 sec, which is statically insignificant compared to training time (2.73 sec). The trade-off lies at the server side, but this can still be scalable since server-side computations can be highly optimized, for example, using careful orchestrated parallelize computation and/or distributed systems.

% \subsection{Communication Analysis}
\textbf{Communication Analysis}: Communication cost is measured by file size (binary/json). \autoref{tab:key-sizes} list key sizes shared once at the start of the protocol by the key dealer.
% \vskip -0.15in
\begin{table}[h!]
\caption{Memory sizes of cryptographic keys}
\label{tab:key-sizes}
% \resizebox{0.49\textwidth}{!}
% {
\centering
\begin{tabular}{|l|r|}
\hline
\textbf{Keys} & \textbf{Size} \\
\hline
pk (public key) & 29.4 MB \\
re (relinearization key) & 352.3 MB \\
rot (rotation key) & 11.27 GB \\
sk (secret key) & 14.7 MB \\
symmetric\_key & 512 Bytes \\
HE Encrypted Symmetric Key & 738.5 MB \\
\hline
\end{tabular}%}
% \vskip -0.1in
\end{table}

For each client, each linear layer that is encrypted symmetrically takes 524 KB in disk, making the whole 2 linear layer neural network 1.048 MB in total. The HE-averaged model output by the server after the HHE protocol takes 16.8 MB in size (8.4 MB per encrypted layer). After aggregation, the server will send the HE encrypted model to each client to decrypt. Hence, for each client, the total communication cost for each round is $1.048 + 16.8 = 17.848\ \text{MB}$. And for 3 clients, the total communication cost for one FL round under HHE is $17.848 * 3 = 53.544\ \text{MB}$. We can see that the majority of communication cost is incurred by the downloading step of the average HE model from the server, while the symmetric model that's needed to be uploaded by each client is 16 times less heavy in size. 

\subsection{Comparison with Plaintext and CKKS-based HE}

% \subsubsection{Accuracy Comparisons}
\textbf{Accuracy Comparisons}: Evaluating the plaintext averaged model, the decrypted averaged models under HE and under HHE all give the exact same accuracies on all test sets:
\begin{itemize}[left=0pt, nosep]
    \item Accuracy on test set with all samples: 65.92\%
    \item Accuracy on test set with labels 1, 3, 7: 73.15\%
    \item Accuracy on test set with labels 2, 5, 8: 54.35\%
    \item Accuracy on test set with labels 4, 6, 9: 58.43\%
\end{itemize}

% \subsubsection{Computation Comparisons}
\textbf{Computation Comparisons}: Note that for \flhhe{}, we only account for the computation time of the online phase for both the client and the server, since the values in the offline phase can be precomputed, for example, the values of the offline phase of the server can be computed while the clients do the training, similarly for the clients. On the client side:
\begin{itemize}[left=0pt, nosep]
    \item In the plaintext setting, each client simply trains the model locally without any encryption.
    \item Under HE, clients first decrypt the global model received from the server, train locally, and then re-encrypt the updated model into HE ciphertext before sending it back.
    \item In \flhhe{}, clients also decrypt the average HE-encrypted model downloaded from the server and train locally, but instead of encrypting with HE, they encrypt their updated model using symmetric encryption before sending it back to the server. \autoref{tab:fl-compute-times} shows client computation time.
\end{itemize}
% \vskip -0.15in

\begin{table}[h!]
\centering
\caption{Computation Time for Different FL Settings (unit: sec)}
\label{tab:fl-compute-times}
\begin{minipage}[t]{0.48\textwidth}
\centering
\textbf{Client Computation}
\vspace{0.5em}
\begin{tabular}{|l|c|c|c|}
\hline
\textbf{Activity} & \textbf{Plain} & \textbf{HE} & \textbf{\flhhe{}} \\
\hline
Decrypt model         & N/A       & 0.07     & 0.07 \\
Local training        & 2.73      & 2.73     & 2.73 \\
Encrypt model         & N/A       & 0.21     & 0.008 \\
\textbf{Total}        & 2.73      & 3.01     & 2.808 \\
\hline
\end{tabular}
\end{minipage}
\hfill
\begin{minipage}[t]{0.48\textwidth}
\centering
\textbf{Server Computation}
\vspace{0.5em}
\begin{tabular}{|l|c|c|c|}
\hline
\textbf{Activity} & \textbf{Plain} & \textbf{HE} & \textbf{\flhhe{}} \\
\hline
Transcipher        & N/A       & N/A     & 6.75 \\
Aggregation        & 0.00005   & 0.004   & 0.003 \\
\textbf{Total}     & 0.00005   & 0.004   & 6.75 \\
\hline
\end{tabular}
\end{minipage}
\end{table}

% \begin{table}[h!]
% \centering
% \caption{Client Computation for Different FL Settings (unit: sec)}
% \label{tab:fl-compute-times}
% % \resizebox{0.49\textwidth}{!}
% % {
% \begin{tabular}{|l|c|c|c|}
% \hline
% \textbf{Activity} & \textbf{Plaintext} & \textbf{HE} & \textbf{\flhhe{}} \\
% \hline
% Decrypt global model         & N/A       & 0.07     & 0.07 \\
% \hline
% Local training               & 2.73    & 2.73     & 2.73 \\
% \hline
% Encrypt updated model        & N/A       & 0.21     & 0.008 \\
% \hline
% \textbf{Total Time}          & 2.73     & 3.01      & 2.808 \\
% \hline
% \end{tabular}
% \end{table}
\begin{tcolorbox}[colback=gray!5!white, colframe=gray!75!black, title=Research Question Answers]
\begin{itemize}[left=0pt, nosep]
    \item This result helps us answer our first research question: ``Can we combine FL and HHE to achieve shared trained models without leaking data privacy?''. The answer is yes, \flhhe{} enables collaborative model training with a focus on privacy.
    \item Regarding the second research question, experimental results show us that HHE does not degrade the accuracy of the FL aggregated model compared to the plaintext or HE version.
\end{itemize}
\end{tcolorbox}
On the server side:
\begin{itemize}[left=0pt, nosep]
    \item In the plaintext case, the server performs standard model averaging on the unencrypted models.
    \item For the HE setting, the server directly aggregates the models within HE domain using homomorphic operations.
    \item With \flhhe{}, server transciphers symmetrically encrypted models received from clients into HE ciphertexts and performs model averaging in HE domain.
\end{itemize}
\autoref{tab:fl-compute-times} shows server-side computation time breakdown for one client for one round. We can see that server-side computation of \flhhe{} is dominated transciphering. After the transciphering step, HE aggregation in FL with HE and \flhhe{} are comparable, however, they are still about 70 times slower than aggregation in plaintext.
% \vskip -0.15in
% \begin{table}[h!]
% \centering
% \caption{Server Computation for Different FL Settings (unit: sec)}
% \label{tab:fl-compute-times}
% \begin{tabular}{|l|c|c|c|}
% \hline
% \textbf{Activity} & \textbf{Plaintext} & \textbf{HE} & \textbf{\flhhe{}} \\
% \hline
% Transcipher       & N/A       & N/A         & 6.75 \\
% Model aggregation          & 0.00005    & 0.004         & 0.003 \\
% \hline
% \textbf{Total Time}                     & 0.00005     & 0.004       & 6.75 \\
% \hline
% \end{tabular}
% \end{table}
% \begin{tcolorbox}[colback=blue!5!white, colframe=blue!75!black, title=Research Question Answers]

\begin{tcolorbox}[colback=gray!5!white, colframe=gray!75!black, title=Research Question Answers]
\flhhe{} offers a reduction in client-side computation compared to a full HE implementation, and adds minimal overhead compared to plaintext. The total online computation time for a client in \flhhe{} is 2.808 seconds, compared to 3.01 seconds for HE and 2.73 seconds for plaintext (~\autoref{tab:fl-compute-times}). The critical symmetric encryption step in \flhhe{} takes only 0.008 seconds, compared to 0.21 seconds for HE, which is 26.25 times faster. This is critical when the model scales up in terms of complexity and size.
\end{tcolorbox}

\begin{tcolorbox}[colback=gray!5!white, colframe=gray!75!black, title=Research Question Answers]
\textbf{How much computation is needed on the server to compensate?} The primary trade-off for efficiencies gained by the client is a substantial increase in server-side computation, mainly due to the transciphering step. Server online computation time per client per round in \flhhe{}: 6.75 seconds (~\autoref{tab:fl-compute-times}). This is significantly higher than both the HE setting (0.004 seconds) and the plaintext setting (0.00005 seconds). This increased load is dominated by the "Transcipher" step, which takes 6.75 seconds.
\end{tcolorbox}

% \subsubsection{Communication Comparisons}
\textbf{Communication Comparisons}: \autoref{tab:comm-costs} shows per-client communication cost per round for plaintext, HE, and HHE.
\vskip -0.15in
\begin{table}[h!]
\centering
\caption{Communication cost per client in one round}
\label{tab:comm-costs}
\resizebox{0.7\textwidth}{!}
{
\centering
\begin{tabular}{|l|r|r|r|}
\hline
\textbf{1 round, 1 client} & \textbf{Send (MB)} & \textbf{Receive (MB)} & \textbf{Total (MB)} \\
\hline
Plain & 0.563 & 0.564 & 1.127 \\
HE & 16.8 & 16.8 & 33.6 \\
HHE & 1.048 & 16.8 & 17.848 \\
\hline
\end{tabular}}
\vskip -0.1in
\end{table}

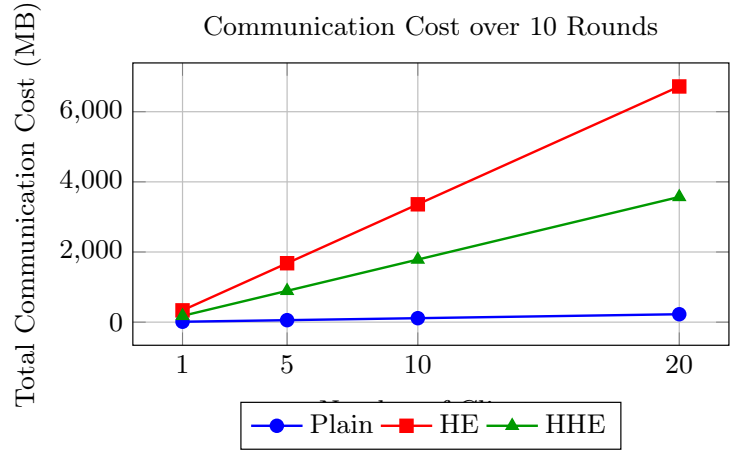
\begin{figure}[h!]
\vskip 0.2in
\begin{center}
\resizebox{0.8\textwidth}{!}
{
\begin{tikzpicture}
\begin{axis}[
    width=0.7\textwidth,
    height=0.4\textwidth,
    xlabel={Number of Clients},
    ylabel={Total Communication Cost (MB)},
    title={Communication Cost over 10 Rounds},
    legend style={at={(0.5,-0.2)},anchor=north,legend columns=-1},
    xtick={1,5,10,20},
    grid=both
]

% Plain: 1.127 MB per round per client
\addplot[
    color=blue,
    mark=*,
    thick
] coordinates {
    (1, 1.127*10)
    (5, 1.127*5*10)
    (10, 1.127*10*10)
    (20, 1.127*20*10)
};
\addlegendentry{Plain}

% HE: 33.6 MB per round per client
\addplot[
    color=red,
    mark=square*,
    thick
] coordinates {
    (1, 33.6*10)
    (5, 33.6*5*10)
    (10, 33.6*10*10)
    (20, 33.6*20*10)
};
\addlegendentry{HE}

% HHE: 17.848 MB per round per client
\addplot[
    color=green!60!black,
    mark=triangle*,
    thick
] coordinates {
    (1, 17.848*10)
    (5, 17.848*5*10)
    (10, 17.848*10*10)
    (20, 17.848*20*10)
};
\addlegendentry{HHE}

\end{axis}
\end{tikzpicture}}
\end{center}
\caption{Communication cost vs. number of clients}
\label{fig:comm-cost}
\vskip -0.1in
\end{figure}
\begin{tcolorbox}[colback=gray!5!white, colframe=gray!75!black, title=Research Question Answers]
\textbf{Client Communication Savings:} \flhhe{} reduces client upload costs.
\begin{itemize}[left=0pt, nosep]
    \item Upload Cost: Clients using \flhhe{} send 1.048 MB, which is 16 times less than the 16.8 MB required with HE.
    \item Total Cost: The total communication cost per client per round for \flhhe{} is 17.848 MB (1.048 MB send + 16.8 MB receive), which is nearly half that of the HE scheme (33.6 MB total).
    \item Scalability: As shown in~\autoref{fig:comm-cost}, HHE's total communication cost scales better than HE as the number of clients increases, which is important in large-scale production settings.
\end{itemize}
\end{tcolorbox}
Overall, while the plaintext version is much cheaper in terms of communication (16 times less than HHE, and 30 times less than HE), it offers no privacy data protection. In contrast, for a client in an FL training workflow, using HHE can save a lot of upload cost compared to HE (16 times less) by sending a symmetric encrypted model instead of an HE encrypted one. Nevertheless, since the FL server returns the same HE-encrypted aggregated model in both cases, the total communication cost of \flhhe{} for each client is reduced only 2 times, compared to HE. However, consider the fact that in real-world FL deployments with many clients participating across many rounds, this reduction can lead to substantial overall bandwidth savings(see~\autoref{fig:comm-cost}).

\section{Generalizable Insights about Responsible Application of Machine Learning in Healthcare}

This work demonstrates that combining lightweight encryption with FL can protect client data without sacrificing model performance. Key insights include:

\begin{itemize}
    \item \textbf{Client-side efficiency:} Reducing client-side cost is crucial for deployment in resource-constrained healthcare settings.
    \item \textbf{Privacy-utility trade-offs:} Lightweight encryption at the client combined with secure processing on the server offers a good balance between privacy and performance.
    \item \textbf{Scalable privacy-preserving FL:} Communication-efficient approaches are essential for real-world healthcare applications.
\end{itemize}

\begin{remark}
Although we did not use a healthcare dataset, the proposed method can be easily extended to such settings and is applicable to any healthcare dataset, supporting responsible and secure ML in healthcare.
\end{remark}

\noindent \textbf{Open Science and Reproducible Research:} 
%\label{subsec:openscience}
To support open science and reproducible research, source code used for the evaluations is publicly available\footnote{\url{https://github.com/khoaguin/flhhe}}.

\section{Conclusion}
\label{sec:conclusion}
In this work, we investigated the integration of HHE with FL, termed \flhhe{}, to address the critical challenges of data privacy and communication overhead in decentralized model training. The protocol uses symmetric encryption on the client-side and HE on the server-side, with a transciphering step to convert between them. This approach is designed to prevent raw data leakage, offering privacy benefits over plaintext FL, and is compared against a full HE-based approach in terms of performance. Experimental results demonstrate that \flhhe{} enables collaborative model training without compromising accuracy, offering enhanced privacy and substantially reduced client communication. While it increases server-side computation, the benefits for client resource efficiency position \flhhe{} as a promising pathway towards more practical, scalable, and secure FL systems. Future work may optimize client downloading cost, improve server-side transciphering and aggregation, or explore real-world applications of \flhhe{}.

\begin{credits}
\subsubsection{\ackname} This work was funded by the HARPOCRATES EU research project (No. 101069535).
\end{credits}
\bibliographystyle{splncs04}
\bibliography{mybibliography}

\end{document}